\def\Title{	Cyclicity of non-associative products on D-branes	}
\def\Abstract{
        The non-commutative geometry of deformation quantization appears in 
        string theory through the effect of a $B$-field background on
        the dynamics of D-branes in the topological limit.
        For arbitrary backgrounds, associativity of the star product is lost, 
        but only cyclicity is necessary for a description of the effective 
        action in terms of a generalized product.
        In previous work we showed that this property indeed emerges for a 
        non-associative product that we extracted from open string 
        amplitudes in curved background fields. In the present note we
        extend our investigation through second order in a complete 
        derivative expansion. We establish cyclicity with respect to 
        the Born--Infeld measure and find a logarithmic correction that 
        modifies the Kontsevich formula in an arbitrary background 
        satisfying the generalized Maxwell equation. This equation is 
        the physical equivalent of a divergence-free $\Th$, which is 
        required for cyclicity already in the associative case.
}

\documentclass[11pt]{article} 

\def\ifundefined#1{\expandafter\ifx\csname#1\endcsname\relax}
\ifundefined{draftmode} {} \else        
   \newcount\HOUR  \HOUR=\the\time \divide\HOUR by 60    \multiply \HOUR by 60
   \newcount\MIN   \MIN=\the\time  \advance\MIN by -\HOUR   \divide\HOUR by 60
   \ifnum   \MIN>9  \def\printTIME{{\it\the\HOUR\,:\,\the\MIN}}
         \else   \def\printTIME{{\it\the\HOUR\,:\,0\the\MIN}} \fi 
   
\fi

\ifundefined{draftmode} {} \else \usepackage[notref,notcite]{showkeys} \fi

\usepackage{latexsym,amsmath,amssymb,cite}

\newcommand {\Ii} {\mathrm{i}}
\newcommand {\Th} {\Theta}
\newcommand {\ud} {\mathrm{d}}
\newcommand {\cF} {{\cal F}}

\newcommand {\cD}{{\cal D}}

\newcommand {\cO}{{\cal O}}

\newcommand {\back}{\!\!\!\!\!\!\!\!\!\!\!\!}
\newcommand {\Back}{\!\!\!\!\!}

\let\a=\alpha       \let\g=\gamma   \let\d=\delta   
           
   \let\l=\lambda  \let\m=\mu      
\let\n=\nu            \let\p=\pi      \let\r=\rho     \let\s=\sigma 
               
            \let\O=\Omega    
      \let\Th=\Theta       
\def\BE{\begin{equation}}      \def\EE{\end{equation}}      \let\6=\partial
\def\BEA{\begin{eqnarray}}   \def\EEA{\end{eqnarray}}

\def\fnote#1#2{\begingroup\def\thefootnote{#1}\footnote{#2}
                \addtocounter{footnote}{-1}\endgroup}

\begin{document} \baselineskip=16pt

\thispagestyle{empty}

\begin{flushright}
TUW--03--33\\
ITP--UH--32/03\\
CERN-TH/2003-280\\
hep-th/0312043
\end{flushright}

\vspace*{3mm}

\begin{center}{\LARGE {\bf 	\Title		}}
\end{center}

\vspace*{5truemm}

\begin{center}
{\large Manfred Herbst,\fnote{\#}{e-mail: Manfred.Herbst@cern.ch}
        Alexander Kling,\fnote{$\,\Box$\,}{e-mail: kling@itp.uni-hannover.de}
        Maximilian Kreuzer\fnote{\,*\,}{e-mail: kreuzer@hep.itp.tuwien.ac.at}
}
\end{center}

{\sl 
\begin{center}
$^{\#}$ Theory Division, CERN\\
CH-1211, Geneva 23, Switzerland\\ 
{$^{\,\Box}$\,}Institut f\"ur Theoretische Physik, Universit\"at Hannover,\\
Appelstra\ss e 2, D-30167 Hannover, Germany\\
$^{*}$Institut f\"ur Theoretische Physik, Technische Universit\"at Wien,\\
Wiedner Hauptstra\ss e 8-10, A-1040 Vienna, Austria 

\end{center}
}

\vspace*{2mm}

\begin{abstract}        	\normalsize	\Abstract
\ifundefined{comments}\else\comments\fi
\end{abstract}
\vfill


\clearpage
\setcounter{page}{1}

\section{Introduction}
\label{sec:intro}

In a seminal paper, M. Kontsevich \cite{Kontsevich:1997vb} gave an explicit 
formula for the deformation quantization of a Poisson structure on 
$\mathbb R^n$ in terms of a formal power series and established the global 
existence on arbitrary Poisson manifolds using formal geometry.
By definition, a deformation quantization is an associative deformation
of the commutative product that is proportional to the Poisson bracket 
$\{f,g\}_\Th=\Th^{\m\n}\6_\m f\6_\n g$ to first order in $\Th$. A symmetric
part of $\Th$ would be a Hochschild cocycle, which can be removed by a gauge 
transformation \cite{Kontsevich:1997vb,Sternheimer:1998yg}. 
The Jacobi identity of the Poisson bracket is equivalent to a vanishing 
Schouten bracket $[\Th,\Th]=0$, which, in turn, is necessary for the 
existence of an associative deformation.

Much work on deformation quantization was stimulated by the 
observation that non-commutative geometry arises in open string theory 
        \cite{Schomerus:1999ug,Ardalan:1999ce,Seiberg:1999vs}. 
The case of a constant background $B$-field was shown to lead to a 
non-commutative product of functions on the world volume of a D-brane, 
which turned out to be given by the Moyal--Weyl formula. 
Cattaneo and Felder \cite{Cattaneo:2000fm,Cattaneo:2001} then gave a physical 
derivation of Kontsevich's formula in terms of a path integral quantization 
of a 
Poisson sigma model \cite{Schaller:1994es}, which corresponds to an open 
string theory in a certain topological limit.  

More recently the situation of open strings in curved backgrounds was 
considered and it was shown  
that the resulting \emph{non-associative} deformation coincides
with Kontsevich's expression at first order of a derivative expansion
\cite{Cornalba:2002sm,Herbst:2001ai,Herbst:2002}.
In \cite{Herbst:2001ai} we argued that a non-vanishing field strength
$H=dB$ of the 2-form $B$-field is incompatible with a topological limit of
Einstein's equations. The dependence on the metric $g_{\m\n}$ 
therefore should not be ignored. The background fields actually include 
a gauge connection 1-form $A$ that lives on the brane in addition to the 
bulk fields $B_{\m\n}$ and $g_{\m\n}$. The equations of motion, however, can 
only depend on the gauge-invariant field strength $H=dB$ and 
$\cF=B+(2\p\a')dA$ of $B$ and $A$. 

Since D-branes can be embedded at arbitrary codimension we expect that only
the variational equation for the gauge field plays a role for the 
non-commutative dynamics while the bulk field backgrounds $B$ and $g$ should 
remain unconstrained. The antisymmetric non-commutativity parameter 
$\Th^{\m\n}$ and the ``open string metric'' $G^{\m\n}$ are related to
these fields by the matrix inversion $G+\Th=(g+\cF)^{-1}$. The generalized 
Maxwell equation,
$G^{\r\s}D_\r\cF_{\s\m}-\frac12\Th^{\r\s}H_{\r\s}{}^\l\cF_{\l\m}=0$, which 
comes from the variation of the Born--Infeld measure $\sqrt{\det(g+\cF)}$ 
with respect to the gauge connection $A$, can thus be recast into the form
\BE     \6_\m(\sqrt{\det(g+\cF)}\Th^{\m\n})=0.  \label{eq:eom}
\end{equation}
For the resulting non-associative product 
\cite{Cornalba:2002sm,Herbst:2001ai}, 
we showed that, to first derivative order, 
\begin{itemize} \vspace{-9pt}
\item   the integrated product of two functions reduces to 
        the ordinary product 
        and that \vspace{-5pt}
\item   the integrated expression for the associator of three functions 
        vanishes
\end{itemize}           \vspace{-9pt}
up to surface terms for the Born--Infeld measure, if the generalized Maxwell
equations (\ref{eq:eom}) are imposed on the background gauge field
\cite{Herbst:2001ai}. This property is called
cyclicity. It is the purpose of the present note to confirm that
cyclicity in the above sense can be extended at least through second order
in the derivative expansion.

The topological limit corresponds to the situation where the metric is 
much smaller than all eigenvalues of $\cF$ so that $\Th\approx\cF^{-1}$.
A vanishing field strength $H=dB=d\cF=0$ (on the D-brane) thus becomes 
equivalent to the Poisson condition $[\Th,\Th]=0$ and the Born--Infeld 
measure reduces to the Liouville measure for the symplectic structure $\Th$.
If we then drop the condition that $\Th$ be invertible and consider
arbitrary Poisson structures the Kontsevich formula still defines a 
deformation quantization, but the natural measure is lost. In that 
context a measure $\O$ has to be introduced as an independent object
\cite{Schoikhet:1999}.
Notably, Felder and Shoikhet constructed a cyclic
(gauge-equivalent) modification of the Kontsevich product 
for Poisson structures $\Th$ that are divergence-free with respect to a 
measure $\O$, i.e. 
\BE     \int\limits_M \Omega\cdot (f*g)\cdot h =        \label{cyc1}
        \int\limits_M \Omega\cdot (g*h)\cdot f 
\end{equation}
for functions $f,g,h\in C^\infty({\cal M})$ of compact support if 
$\mathop{\rm div}_\O\Th=0$ \cite{Felder:2000nc}. 
Using the identity $g*1 = 1*g = g$ this immediately implies the generalized
Connes--Flato--Sternheimer conjecture \cite{Connes:1992}:
\BE     \int\limits_M \Omega\cdot (f*g) = \int\limits_M \Omega\cdot f\cdot g .
                                                                \label{cyc2}
\end{equation}
In the context of open string theory, there exists a natural measure 
regardless of 
the rank of $\cF$ or of $\Th$, and the divergence condition has the natural
interpretation of a generalized Maxwell equation (\ref{eq:eom}) if $\O$ is 
identified with the Born--Infeld measure $\sqrt{\det(g-\cF)}~d^Dx$.
Moreover, cyclicity (\ref{cyc1}, \ref{cyc2}) of the deformed product 
can be preserved, at least through second derivative order, even in the 
non-associative case. 
We conjecture that this property can be maintained to all orders, but it may
then become necessary to take into account derivative corrections to the
Born--Infeld measure\cite{Okawa:1999cm,Andreev:2001xx,Wyllard:2000qe,Fotopoulos:2001pt,Das:2001xy,Wyllard:2001ye,Pal:2001xp}.

In this note we 
explore the cyclicity property at second derivative order of the background 
fields. Since a diagrammatic calculation along the lines of 
\cite{Herbst:2001ai} would be extremely tedious, we check the consistency of
our proposal with an ansatz. We should, of course, reproduce the topological
limit, which essentially fixes the product up to gauge equivalence. Since
we need the explicit expression for the associator we first include all
Kontsevich-type graphs without loops with arbitrary coefficients. 
Associativity up to terms proportional to the `Jacobiator' 
$J=\frac32[\Th,\Th]$
of the Poisson bracket then fixes all coefficients of the ansatz, except for 
a contribution to the product that is itself proportional to $J$. (Obviously 
such a term is not constrained by associativity, but it can be fixed by a 
symmetry argument.) Thus we recover
the known results of \cite{Penkava:1998xx,Zotov:2001ec} 
and, since we are working in a
derivative expansion, extend them to all orders in
the constant part of the non-commutativity parameter.

The main focus will then lie on the verification of the cyclicity property. 
Using the equations of motion for the background gauge field
and the expression for the associator, cyclicity also fixes the gauge part
of the product. We thus recover the contribution from a loop diagram 
in Kontsevich's expansion with the same coefficient that was explicitly 
calculated in \cite{Dito:2002dr}. 
In addition, we find a new term with a logarithmic derivative of the 
Born--Infeld measure, which restores cyclicity up to terms with at least
three derivatives on the background fields $\Th$ and $G$.

The paper is organized as follows: Section \ref{sec:ETP} contains a
discussion of the physical relevance of the cyclicity property and a
brief review of the results of \cite{Herbst:2001ai}.
In section \ref{sec:asscyc} we present the ansatz for the
non-commutative product and derive the modifications 
that are required by cyclicity. 
We conclude with a discussion of our results.
The evaluation of the associator is outlined in the appendix.

\section{Physical relevance of the cyclicity property}
\label{sec:ETP}

The requirement of a cyclicity property has shown up on a fundamental
level of string theory in several places. In the context of open
string field theory it constitutes a necessary prerequisite
to be able to write down an action which satisfies the BV master
equation~\cite{Gaberdiel:1997ia}. An analogous statement is known for
closed string field theory~\cite{Zwiebach:1992ie} and topological
strings~\cite{Alexandrov:1995kv}. 
In this section we discuss why cyclicity of a non-commutative product
is a desirable property in an effective action arising from open
string theory, regardless of the associativity of the product.
Our arguments will be based on the Lagrangian formalism and the
variational principle of a (space-time) quantum field theory and on
modular invariance of open string theory on the disk.
These considerations are quite general and apply to the full
non-commutative product emerging from string theory.

As is well known, the space-time low energy effective action can be
obtained by
computing string amplitudes; the equations of motion for the string
background fields emerge from calculating the conformal anomaly. 
Both quantities should be related by the variational principle in the
low energy effective theory. 
Turning to the perspective of string theory, the purpose of
introducing a non-commutative product on the world-volume of a D-brane is to 
sum up the effect of the background fields in an elegant way.
We expect that both the action and the equation of motion can be expressed in 
terms of a non-commutative product, which means that the antisymmetric 
background field $\Theta$ should only appear implicitly via the
product. What are the implications of such an assumption? 

To illustrate these considerations we 
pick some interaction term, say $\int \Phi \circ (\Phi \circ \Phi)$. Applying
the variational principle in order to obtain the equations of motion, we  
obviously obtain three terms. 
From SL(2,R) invariance of disk on-shell correlators and from 
the fact that the properties of the product should not depend on the
on-shell condition of the functions (because of the lack of a proper metric 
dependence), we can expect that the trace property of the integral
holds, i.e. that
\begin{eqnarray}
  \label{eq:assocprod}
  \int (f \circ  g) \circ h &=& 
  \int f \circ (g \circ  h) ~,
\end{eqnarray}
as well as $\int f\circ g = \int g\circ f$. 
Then our variation takes
the form $3 \int \delta\Phi \circ (\Phi \circ \Phi)$.%
\footnote{Interactions with higher powers in the fields yield
          a sum over different positionings of brackets.
}
To separate the contribution to the equation of motion we still have to 
remove all derivatives from the variation $\delta\Phi$. Doing this by partial 
integration would produce an explicit $\Theta$ dependence in the
equation of motion. Therefore, we infer that as a building block of a field 
theory the product should obey
\begin{eqnarray}
  \label{eq:ordprod}
  \int f \circ g &=& 
  \int f \cdot g ~.
\end{eqnarray} 
Sticking to our example, we obtain 
$3 \int \delta\Phi \cdot (\Phi \circ \Phi)$, which gives rise to 
$\Phi \circ \Phi$ in the equation of motion. 
Therefore we expect that the
low energy field theory obtained from open string theory contains a 
(generically non-commutative and non-associative) product that
satisfies the cyclicity property (\ref{eq:assocprod}) and (\ref{eq:ordprod}).

In order to see how this works at first derivative order,
and as a warm-up for the calculation in the next section,
we briefly review the non-commutative product found in
\cite{Herbst:2001ai}. 
It was obtained from the computation of off-shell correlators of an 
open string sigma model with arbitrary, massless, on-shell
background fields 
apart from the dilaton, which was set to zero. In the bulk these are
the background metric $g$ and the antisymmetric $B$-field and at the
boundary it is the gauge field $A$. 
The product found in \cite{Herbst:2001ai} to all orders in 
the non-commutative parameter $\Theta$ and to first derivative order
in the background fields is
\begin{eqnarray}
  \label{eq:NCP}
  f(x) \; \circ \; g(x) \;=\; f * g 
  \Back&-&\Back \frac 1{12}
  \Theta^{\mu\rho} \partial_\rho \Theta^{\nu\sigma} \;
  \Bigl(\partial_\mu \partial_\nu f \;* \partial_\sigma g +
  \partial_\sigma f \;* \partial_\mu \partial_\nu g \Bigr)
  + \cO\bigl((\partial\Theta)^2,\partial^2\Theta\bigr) ,
\end{eqnarray}
where `$*$' denotes the Moyal contribution to the product, 
\begin{equation}
  \label{eq:Moyalpart}
  f(x) * g(x) = e^{\frac i2\Theta^{\mu\nu}(x)
                \partial_{u^\mu}\partial_{v^\nu}}
                f(u)~g(v)\bigr|_{u=v=x} .
\end{equation}
Although formally the same, this represents a non-associative version of 
Kontsevich's star product formula since $\Theta$ is not assumed to define a 
Poisson structure. Associativity of (\ref{eq:NCP}) is violated by terms 
proportional to the Jacobi identity 
\begin{eqnarray}
  \label{eq:nonassoc1O}
  (f \circ g) \circ h - f \circ (g \circ h) \; = 
       \frac{1}{6} \bigl( 
         \Theta^{\mu\sigma}\partial_\sigma \Theta^{\nu\rho}
        +\Theta^{\nu\sigma}\partial_\sigma \Theta^{\rho\mu}
        +\Theta^{\rho\sigma}\partial_\sigma \Theta^{\mu\nu} 
        \bigr)  
  [\partial_\mu f * \partial_\nu g * \partial_\rho h] +\cO(\partial^2) ,
\end{eqnarray}
where we introduced the abbreviation
\begin{equation}
  \label{eq:bracket}
  [f * g * h] = e^{\frac i2\Theta^{\mu\nu}(x)
                   (\partial_{u^\mu}\partial_{v^\nu} +
                    \partial_{u^\mu}\partial_{w^\nu} +
                    \partial_{v^\mu}\partial_{w^\nu})} 
                f(u)~g(v)~h(w)\bigr|_{u=v=w=x},
\end{equation}
which denotes the Moyal-type triple product with all terms containing
derivatives on $\Th$ removed. Imposing the generalized Maxwell equation 
(\ref{eq:eom}), it was
shown in \cite{Herbst:2001ai} that the product (\ref{eq:NCP})
satisfies the cyclicity relations (\ref{eq:assocprod}) and
(\ref{eq:ordprod}) to first derivative order of the background
fields. The relevant integration measure is given by the Born--Infeld measure
$\int =\int \ud^{D} x \sqrt {\det(g\!\!-\!\!\cF)}$, which
arises from the vacuum amplitude. This result confirms our general
arguments above and motivates us to look at further derivative
corrections to the product. 

Before we go on to the next section and consider the second derivative
order, we want to make a comment concerning the Moyal-type triple
product $[f * g * h]$. It differs from both $f*(g*h)$ and $(f*g)*h$.
Expressions like (\ref{eq:bracket}) are useful for the evaluation of
derivative expansions, since they automatically keep all orders in
the undifferentiated $\Th$. We should keep in mind, however, that
we actually work with a double expansion because already the
Moyal-type contributions have to be understood as formal power series.
We do not use the conventional $\hbar$ to indicate this fact because
our derivative expansion is a formal power series in two variables,
controlling the number of $\Th$'s and the number of derivatives
acting on them, respectively. Our formulas keep terms of arbitrary
order in the first parameter and we drop all terms that are cubic in
the second one.

\section{The non-associative product at second derivative order}
\label{sec:asscyc}

In order to check for the cyclicity of the `$\circ$' product, we first need to 
evaluate the associator. For this purpose it is sufficient to drop all 
Hochschild coboundaries, i.e. all terms that can be gauged away by a 
transformation 
\BE     f\circ g\to \cD^{-1}(\cD f\circ \cD g)
        , ~~~~ \cD=1+A^{\m\n}\6_\m\6_\n+\ldots	\label{GaugeTrafo}
\end{equation}
where $\cD$ is some formally invertible differential operator. In particular, 
contributions to $f\circ g$ of the form $X^{\m\n}\6_\m f\6_\n g$ with
symmetric $X$ can be gauged away with $A^{\m\n}=-X^{\m\n}$.
We thus start with an ansatz for the product that contains expressions with
arbitrary coefficients for all Kontsevich-type graphs with two derivatives 
acting on $\Th$, as displayed in fig. 1.

\begin{figure}
                        \def\putc#1)#2{\put#1){\makebox(0,0)[c]{#2}}}
\begin{center}                          \unitlength=2pt\thicklines
\begin{picture}(100,20)(-50,-10)\large
\put(-90,5){\putc(10,0){$\Th$}\putc(-10,0){$\Th$}\putc(0,-15){(A)}
        \put(-7,-1.5){\vector(1,0){14}} \put(7,1.5){\vector(-1,0){14}}  }
\put(-45,5){\putc(12,4){$\Th$}\putc(-12,4){$\Th$}\putc(0,-15){(B)}
        \put(-9,2.5){\vector(1,-1){6}}  \put(9,2.5){\vector(-1,-1){6}}  
        \putc(0,-5){$\Th$}      }
\put(0,5){\putc(15,0){$\Th$}\putc(-15,0){$\Th$}\putc(0,-15){(C)}
        \put(-12,0){\vector(1,0){9}}    \put(3,0){\vector(1,0){9}}      
        \putc(0,0){$\Th$}       }
\put(45,5){\putc(4,4){$\Th$}\putc(-4,4){$\Th$}\putc(0,-15){(D)}
        \put(-6.5,2.5){\vector(-1,-1){5}}\put(6.5,2.5){\vector(1,-1){5}}
        \putc(13,-5){$\Th$}\putc(-13,-5){$\Th$} }
\put(90,5){\putc(12,-5){$\Th$}\putc(-12,-5){$\Th$}\putc(0,-15){(E)}
        \put(-3.5,2.5){\vector(-1,-1){6}} \put(3.5,2.5){\vector(1,-1){6}}
        \putc(0,4){$\Th$}       }
\end{picture}
\\[7mm] Figure 1: Graphs with two derivatives acting on $\Th$.
\end{center}
\end{figure}

Abbreviating derivatives acting on $f$ and $g$ with subscripts, we obtain
the following contributions to the product from Kontsevich-type graphs
\cite{Kontsevich:1997vb} that contain structures of the
form (A) $\ldots$ (E) in fig.~1:
\BEA\label{eq:Kgraphs}
  f\circ g &\Back =\Back & f * g 
    -\frac{1}{12}\Th^{\mu\gamma}\partial_{\gamma}\Th^{\nu\rho}
       (f_{\mu\nu}*g_{\rho}+f_{\rho}*g_{\mu\nu}) 
        +\frac14\;\6_\d\Th^{\m\g}\6_\g\Th^{\n\d}(A\;f_\m*g_\n)\nonumber 
\\       & & -\frac{\Ii}{8}
  \Th^{\mu\gamma}\Th^{\nu\delta}
     \partial_{\gamma}\partial_{\delta}\Th^{\rho\lambda}
        (B_1 f_{\mu\nu\rho}*g_{\lambda} + B_2 f_{\lambda}*g_{\mu\nu\rho} 
         + B_3 f_{\mu\rho}*g_{\nu\lambda}) 
\\
       & & -\frac{\Ii}{8}
  \Th^{\mu\gamma}\partial_{\gamma}\Th^{\nu\delta}
     \partial_{\delta}\Th^{\rho\lambda}
        (C_1 f_{\mu\rho}*g_{\nu\lambda}+ C_2 f_{\nu\lambda}*g_{\mu\rho}
         + C_3 f_{\mu\nu\rho}*g_{\lambda}+ C_4 f_{\lambda}*g_{\mu\nu\rho}) 
          \nonumber \\
       & & +\frac{1}{16}(\Th^{\mu\gamma}\partial_{\gamma}\Th^{\rho\lambda})
                        (\Th^{\nu\delta}\partial_{\delta}\Th^{\sigma\tau})
         (D_1 f_{\mu\rho\nu\sigma}*g_{\lambda\tau} 
          + D_2 f_{\mu\rho\tau}*g_{\lambda\nu\sigma}
          + D_3 f_{\lambda\tau}*g_{\mu\rho\nu\sigma}). \nonumber
\EEA
There is no contribution from (E) because the only possible term
$\Th^{\delta\gamma}\partial_{\gamma}\Th^{\mu\nu}
\partial_{\delta}\Th^{\rho\lambda}f_{\mu\rho}*g_{\nu\lambda}$
vanishes due to symmetry of $\m\r$ and $\n\l$ and antisymmetry of $\Th$.
The contribution of graph (A) can be gauged away and
hence does not contribute to the associator
$(f\circ g)\circ h-f\circ (g\circ h)$. Nevertheless, it does
contribute to the Kontsevich product with a coefficient $A=\frac16$,
which is exactly what we will need for cyclicity. 

The evaluation of the associator in the appendix shows that 
consistency with the topological limit fixes 
\BE     B_1=-B_2=\frac16,~~~B_3=0, ~~~~~~~ C_2-C_1=\frac13,~~~C_3=C_4=0, 
        ~~~~~~~ D_1=2D_2=D_3=\frac1{18}.\label{eq:coeff}
\end{equation}
The ambiguity $C_1\to C_1-C_J$ and $C_2\to C_2-C_J$ had to be expected 
because a contribution of the form 
$\frac i8C_JJ^{\m\n\d}\6_\d\Th^{\r\s}f_{\m\r}*g_{\n\s}$ with
\BE     J^{\m\n\d}={\textstyle\frac32}[\Th,\Th]^{\m\n\d}=
        \Th^{\m\g}\6_\g\Th^{\n\d}+\Th^{\n\g}\6_\g\Th^{\d\m}
        +\Th^{\d\g}\6_\g\Th^{\m\n}      \label{eq:jacobiator}
\end{equation}
generates that shift and vanishes for $J=0$
(the last term in (\ref{eq:jacobiator}) yields a contribution of the form (E) 
that vanishes identically).
The Kontsevich formula inherits invariance under the parity transformation
exchanging $f$ and $g$ and the sign of $\Th$ from string theory
via its topological limit. 
This symmetry exchanges 
$C_1$ with $-C_2$, $B_1$ with $-B_2$, $D_1$ with $D_3$ and leaves all other 
terms invariant, which implies that the
appropriate value is $C_1=-C_2=-\frac16$.
For the sake of generality we 
will, however, keep the $C_2$ dependence in the following expressions. 
The resulting product reads
\begin{eqnarray}
 \label{eq:product} 
 f\circ g &\Back =\Back & f * g 
    -\frac{1}{12}\Th^{\mu\gamma}\partial_{\gamma}\Th^{\nu\rho}
       \Bigl((f_{\mu\nu}*g_{\rho}+f_{\rho}*g_{\mu\nu}\Bigr) \nonumber 
    +\frac14\;\6_\d\Th^{\m\g}\6_\g\Th^{\n\d}~\Bigl(A\;f_\m*g_\n \Bigr)
\\ 
& & -\frac{\Ii}{48}\Th^{\mu\gamma}\Th^{\nu\delta}
     \partial_{\gamma}\partial_{\delta}\Th^{\rho\lambda}
     \Bigl(f_{\mu\nu\rho}*g_{\lambda} -
     f_{\lambda}*g_{\mu\nu\rho}\Bigr) \nonumber 
\\
& & -\frac{\Ii}{8} \Th^{\mu\gamma}\partial_{\gamma}\Th^{\nu\delta}
  \partial_{\delta}\Th^{\rho\lambda}
  \Bigl(\Bigl(C_2-\frac{1}{3}\Bigr)\ f_{\mu\rho}*g_{\nu\lambda}
   + C_2\ f_{\nu\lambda}*g_{\mu\rho}\Bigr) \nonumber
\\
& & +\frac{1}{2}\frac{1}{12^2}
      (\Th^{\mu\gamma}\partial_{\gamma}\Th^{\rho\lambda})
      (\Th^{\nu\delta}\partial_{\delta}\Th^{\sigma\tau})
      \Bigl((f_{\mu\rho\nu\sigma}*g_{\lambda\tau} 
       +2 f_{\mu\rho\tau}*g_{\lambda\nu\sigma}
       + f_{\lambda\tau}*g_{\mu\rho\nu\sigma}\Bigr),
\end{eqnarray}
and for the associator we obtain
\begin{eqnarray}
  \label{eq:nonassoc}
  (f \circ g)\circ h &\Back -\Back &  f \circ (g \circ h) = 
  \frac{1}{6} J^{\mu\nu\rho}[f_{\mu}*g_{\nu}*h_{\rho}] \nonumber 
\\  & \Back \back & ~~~
  +2\,\Bigl(\frac{1}{12}\Bigr)^2
  (\Th^{\mu\gamma}\partial_{\gamma}\Th^{\rho\lambda})
        J^{\nu\sigma\tau} ~\Bigl( [f_{\mu\rho\nu}*g_{\tau}*h_{\lambda\sigma}]
    + [f_{\mu\rho\nu}*g_{\lambda\tau}*h_{\sigma}] \nonumber
\\    & \back  \back & ~~~~~~~~~~~~~~~~~
    + [f_{\nu\lambda}*g_{\mu\rho\tau}*h_{\sigma}] 
    + [f_{\nu}*g_{\mu\rho\tau}*h_{\lambda\sigma}] \nonumber
    +2\, [f_{\mu\nu}*g_{\rho\tau}*h_{\lambda\sigma}]
\\    & \back  \back & ~~~~~~~~~~~~~~~~~ 
    +2\, [f_{\lambda\nu}*g_{\rho\tau}*h_{\mu\sigma}] 
    + [f_{\nu}*g_{\lambda\tau}*h_{\mu\rho\sigma}] 
    + [f_{\nu\lambda}*g_{\tau}*h_{\mu\rho\sigma}]\Bigr)  \nonumber
\\  & \Back  \Back & ~~~
  +\frac{\Ii}{24}J^{\mu\nu\delta}\partial_{\delta}\Theta^{\rho\lambda} 
  ~\Bigl(3C_2[f_{\mu\rho}*g_{\nu}*h_{\lambda}]
  +(3C_2-1)[f_{\mu\rho}*g_{\lambda}*h_{\nu}] \nonumber
\\  & \Back  \Back & ~~~~~~~~~~~~~~~~~
  -3C_2[f_{\mu}*g_{\rho}*h_{\nu\lambda}]
  -(3C_2-1)[f_{\rho}*g_{\mu}*h_{\nu\lambda}] \nonumber
\\  & \Back  \Back & ~~~~~~~~~~~~~~~~
  +[f_{\mu}*g_{\nu\rho}*h_{\lambda}]
  +[f_{\lambda}*g_{\nu\rho}*h_{\mu}] \Bigr)~ \nonumber
\\  & \Back  \Back & ~~~
   +\frac{\Ii}{24}\Theta^{\mu\gamma}\partial_{\gamma}J^{\nu\rho\lambda} 
   ~\Bigl([f_{\mu\nu}*g_{\rho}*h_{\lambda}]-
   [f_{\rho}*g_{\lambda}*h_{\mu\nu}]\Bigr),
\end{eqnarray}
where each term contains the Jacobiator (\ref{eq:jacobiator}) as required
by consistency with the topological limit.

\noindent\emph{Associativity up to surface terms}

We will now check relation (\ref{eq:assocprod}) for the product 
(\ref{eq:product}) to second derivative order. To this end we integrate the 
associator (\ref{eq:nonassoc}) with the measure, $\sqrt{\det(g-\cF)}$, and 
take the equations of motion for the background fields into account. We will 
find that lines 2 -- 4 of (\ref{eq:nonassoc}) vanish by themselves. The 
same holds for lines 5 and 6. The first line can be pushed to second 
derivative order and cancels the last two lines.

We start with the easiest piece, the terms proportional to 
$(\Theta\partial\Theta)J$. In fact, these can all 
be pushed into the third derivative order by partially integrating one of the 
derivatives contracted with $J$, e.g.
\begin{eqnarray}
  \label{eq:NA2Ord1}
  \int d^Dx \sqrt{\det (g - \cF)}&\Back \Back & 
    \Theta^{\mu\gamma}\partial_{\gamma}\Theta^{\rho\lambda}J^{\nu\sigma\tau} 
    [f_{\mu\rho\nu}*g_{\tau}*h_{\lambda\sigma}] = s.t. \nonumber
\\ &\Back \Back &
      -\int d^Dx \partial_{\nu}(\sqrt{\det (g - \cF)}J^{\nu\sigma\tau})
      \Theta^{\mu\gamma}\partial_{\gamma}\Theta^{\rho\lambda} 
        [f_{\mu\rho}*g_{\tau}*h_{\lambda\sigma}]  \nonumber
\\ &\Back \Back &
      -\int d^Dx (\sqrt{\det (g - \cF)}J^{\nu\sigma\tau})
        \partial_{\nu}(\Theta^{\mu\gamma}\partial_{\gamma}\Theta^{\rho\lambda})
        [f_{\mu\rho}*g_{\tau}*h_{\lambda\sigma}] \nonumber
\\ &\Back \Back &
      -\int d^Dx (\sqrt{\det (g - \cF)}J^{\nu\sigma\tau})
        \Theta^{\mu\gamma}\partial_{\gamma}\Theta^{\rho\lambda}
        \partial_{\nu}^*[f_{\mu\rho}*g_{\tau}*h_{\lambda\sigma}] \nonumber
\\ &\Back \approx \Back & ~~0~ + ~\cO(\partial^3),
\end{eqnarray}
where the derivative $\partial_{\nu}^*$ acts only on the `stars' in the 
product $[f_{\mu\rho}*g_{\tau}*h_{\lambda\sigma}]$, since $J$ is totally 
antisymmetric. In a similar way the two lines containing the constant $C_2$ 
vanish by partially integrating twice.

The remaining second derivative terms in (\ref{eq:nonassoc}) are mixed
up with the first derivative order. Therefore let us concentrate on
the latter and rewrite it as
\begin{eqnarray}
  \label{eq:nonass1ord}
&& \frac{1}{6}\int d^Dx \sqrt{\det (g - \cF)}
          J^{\mu\nu\rho}[f_{\mu}*g_{\nu}*h_{\rho}] 
\\  && ~~~~~~~~~~ = 
s.t.-\frac{1}{6}\int d^Dx\,\partial_{\mu}(\sqrt{\det (g -\cF)}J^{\mu\nu\rho})
 [f*g_{\nu}*h_{\rho}]  \nonumber
\\ &&
 -\frac{\Ii}{12}\int d^Dx \sqrt{\det (g - \cF)}J^{\mu\nu\rho}\partial_{\mu}
 \Theta^{\alpha\beta}([f_{\alpha}*g_{\nu\beta}*h_{\rho}] 
                    + [f_{\alpha}*g_{\nu}*h_{\rho\beta}]
                    + [f*g_{\nu\alpha}*h_{\rho\beta}]). \nonumber
\end{eqnarray}
The second line of eq. (\ref{eq:nonass1ord}) can be shown to vanish 
because of the relation
\begin{equation}
  \label{eq:Jeom}
  \partial_{\mu}(\sqrt{\det (g -\cF)}J^{\mu\nu\rho}) \approx 0,
\end{equation}
which holds by way of the equations of motion of the background field
(\ref{eq:eom}). This can be seen as follows. 

Expanding the Jacobiator we find
\begin{eqnarray}
  \label{eq:eom1}
  \partial_{\mu}\bigl(\sqrt{\det (g - \cF)}J^{\mu\nu\rho}\bigr) &\approx&
  \bigl(\partial_{\mu}\sqrt{\det (g - \cF)}\bigr)~
  (\Theta^{\nu\gamma}\partial_{\gamma}\Theta^{\rho\mu}
  +\Theta^{\rho\gamma}\partial_{\gamma}\Theta^{\mu\nu}) \nonumber
\\ &+&
  \sqrt{\det (g - \cF)}~
  (\partial_{\mu}\Theta^{\nu\gamma}\partial_{\gamma}\Theta^{\rho\mu}
  +\partial_{\mu}\Theta^{\rho\gamma}\partial_{\gamma}\Theta^{\mu\nu})\nonumber
\\ &+&
  \sqrt{\det (g - \cF)}~
  (\Theta^{\nu\gamma}\partial_{\mu}\partial_{\gamma}\Theta^{\rho\mu}
  +\Theta^{\rho\gamma}\partial_{\mu}\partial_{\gamma}\Theta^{\mu\nu}),
\end{eqnarray}
where the second line vanishes identically because of the antisymmetry of 
$\Theta$. Next we exchange the partial derivatives in the last line of 
(\ref{eq:eom1}) and use the background field equation (\ref{eq:eom}),
obtaining 
\begin{eqnarray}
  \label{eq:eom2}
  \partial_{\mu}\bigl(\sqrt{\det (g - \cF)}J^{\mu\nu\rho}\bigr) 
  & \back \approx \back &
  \bigl(\partial_{\mu}\sqrt{\det (g - \cF)}\bigr)~
  (\Theta^{\nu\gamma}\partial_{\gamma}\Theta^{\rho\mu}
  +\Theta^{\rho\gamma}\partial_{\gamma}\Theta^{\mu\nu})
\\ & &
  -\sqrt{\det (g - \cF)}~\biggl(\Theta^{\nu\gamma}\partial_{\gamma}
     \Bigl(\partial_{\mu}\sqrt{\det (g - \cF)}~\Theta^{\rho\mu}
      \frac{1}{\sqrt{\det (g - \cF)}}\Bigr) \nonumber
\\ & & \hspace{2.4truecm}
 +~\Theta^{\rho\gamma}\partial_{\gamma}\Bigl(\partial_{\mu}\sqrt{\det(g-\cF)})
      \Theta^{\mu\nu}\frac{1}{\sqrt{\det (g - \cF)}}\Bigr)\biggr).\nonumber
\end{eqnarray}
The terms where the partial derivative in the second and third lines acts on 
the $\Theta$'s cancel the contributions from the first line, while the other 
terms cancel again, owing to the antisymmetry of the $\Theta$'s. Thus we have 
established our claim (\ref{eq:Jeom}), which shows that from
(\ref{eq:nonass1ord}) only the last line 
\begin{eqnarray}
 \label{eq:NA1O}
-\frac{\Ii}{12}\int d^Dx \sqrt{\det (g - \cF)}J^{\mu\nu\rho}\partial_{\mu} 
\Theta^{\alpha\beta}([f_{\alpha}*g_{\nu\beta}*h_{\rho}] 
                    + [f_{\alpha}*g_{\nu}*h_{\rho\beta}]
                    + [f*g_{\nu\alpha}*h_{\rho\beta}])
\end{eqnarray} 
survives. It has to be considered together with other $J\partial\Theta$ 
contributions in (\ref{eq:nonassoc}). 

To this end we try to transform the last line of
(\ref{eq:nonassoc}) into this form. As a first step we rewrite it as
\begin{eqnarray}
  \label{eq:firstord}
  \frac{\Ii}{24}&\Back \Back &\int d^Dx \sqrt{\det (g - \cF)}
  \Theta^{\mu\gamma}\partial_{\gamma}J^{\nu\rho\lambda}
  ([f_{\mu\nu}*g_{\rho}*h_{\lambda}]-[f_{\rho}*g_{\lambda}*h_{\mu\nu}])
  \approx \nonumber
\\ &\Back \Back & 
  \frac{1}{48}\int d^Dx \sqrt{\det (g - \cF)}
  J^{\nu\rho\lambda}\Theta^{\mu\gamma}\partial_{\gamma}\Theta^{\alpha\beta}
  ([f_{\mu\nu\alpha}*g_{\rho\beta}*h_{\lambda}]
  +[f_{\mu\nu\alpha}*g_{\rho}*h_{\lambda\beta}] \nonumber
\\ &\Back \Back & ~~~~~~~~~~~~~~~~~~~~~~~~~~~~~~~~~~~~~~~~~~~~~~~~~
  +[f_{\mu\nu}*g_{\rho\alpha}*h_{\lambda\beta}]
  -[f_{\rho\alpha}*g_{\lambda\beta}*h_{\mu\nu}] \nonumber
\\ &\Back \Back & ~~~~~~~~~~~~~~~~~~~~~~~~~~~~~~~~~~~~~~~~~~~~~~~~~
  -[f_{\rho\alpha}*g_{\lambda}*h_{\mu\nu\beta}]
  -[f_{\rho}*g_{\lambda\alpha}*h_{\mu\nu\beta}]) \nonumber
\\ &\Back \Back &
 -\frac{\Ii}{24}\int d^Dx \sqrt{\det (g - \cF)}
  J^{\nu\rho\lambda}\Theta^{\mu\gamma}
 ([f_{\mu\nu}*g_{\rho\gamma}*h_{\lambda}]
 +[f_{\mu\nu}*g_{\rho}*h_{\lambda\gamma}] \nonumber
\\ &\Back \Back & ~~~~~~~~~~~~~~~~~~~~~~~~~~~~~~~~~~~~~~~~~
 -[f_{\rho\gamma}*g_{\lambda}*h_{\mu\nu}]
 -[f_{\rho}*g_{\lambda\gamma}*h_{\mu\nu}]),
\end{eqnarray}
where only the last expression cannot be written as surface term.
Note that this is of first derivative order. By partially integrating
$\partial_{\nu}$ we obtain 
\begin{eqnarray}
  \label{eq:NA2Ord2}
  \frac{\Ii}{24}&\Back \Back &\int d^Dx \sqrt{\det (g - \cF)}
  \Theta^{\mu\gamma}\partial_{\gamma}J^{\nu\rho\lambda}
  ([f_{\mu\nu}*g_{\rho}*h_{\lambda}]-[f_{\rho}*g_{\lambda}*h_{\mu\nu}])
  \approx \nonumber
\\ &\Back \Back & 
  \frac{\Ii}{24}\int d^Dx \sqrt{\det (g - \cF)}
  J^{\nu\rho\lambda}\partial_{\nu}\Theta^{\mu\gamma}
  ([f_{\mu}*g_{\rho\gamma}*h_{\lambda}]
  +[f_{\mu}*g_{\rho}*h_{\lambda\gamma}] \nonumber
\\ &\Back \Back & ~~~~~~~~~~~~~~~~~~~~~~~~~~~~~~~~~~~~~~~~~
  -[f_{\rho\gamma}*g_{\lambda}*h_{\mu}]
  -[f_{\rho}*g_{\lambda\gamma}*h_{\mu}]) \nonumber
\\ &\Back \Back &
  -\frac{1}{48}\int d^Dx \sqrt{\det (g - \cF)}
  J^{\nu\rho\lambda}\Theta^{\mu\gamma}\partial_{\nu}\Theta^{\alpha\beta}
  ([f_{\mu\alpha}*g_{\rho\gamma\beta}*h_{\lambda}]
  +[f_{\mu\alpha}*g_{\rho\gamma}*h_{\lambda\beta}] \nonumber
\\ &\Back \Back & ~~~~~~~~~~~~~~~~~~~~~~~~~~~~~~~~~~~~~~~~~
  +[f_{\mu}*g_{\rho\gamma\alpha}*h_{\lambda\beta}]
  +[f_{\mu\alpha}*g_{\rho\beta}*h_{\lambda\gamma}] \nonumber
\\ &\Back \Back & ~~~~~~~~~~~~~~~~~~~~~~~~~~~~~~~~~~~~~~~~~
  +[f_{\mu\alpha}*g_{\rho}*h_{\lambda\gamma\beta}]
  +[f_{\mu}*g_{\rho\alpha}*h_{\lambda\gamma\beta}] \nonumber
\\ &\Back \Back & ~~~~~~~~~~~~~~~~~~~~~~~~~~~~~~~~~~~~~~~~~
  -[f_{\rho\gamma\alpha}*g_{\lambda\beta}*h_{\mu}]
  -[f_{\rho\gamma\alpha}*g_{\lambda}*h_{\mu\beta}] \nonumber
\\ &\Back \Back & ~~~~~~~~~~~~~~~~~~~~~~~~~~~~~~~~~~~~~~~~~
  -[f_{\rho\gamma}*g_{\lambda\alpha}*h_{\mu\beta}]
  -[f_{\rho\alpha}*g_{\lambda\gamma\beta}*h_{\mu}] \nonumber
\\ &\Back \Back & ~~~~~~~~~~~~~~~~~~~~~~~~~~~~~~~~~~~~~~~~~
  -[f_{\rho\alpha}*g_{\lambda\gamma}*h_{\mu\beta}]
  -[f_{\rho}*g_{\lambda\gamma\alpha}*h_{\mu\beta}]).
\end{eqnarray}
The last twelve terms in expression (\ref{eq:NA2Ord2}) cancel by partially 
integrating with respect to $\partial_{\gamma}$, modulo higher derivative 
orders. Thus we are left with the four terms 
\begin{eqnarray}
  \label{eq:NA2Ord3}
& &   \frac{\Ii}{24}\int d^Dx \sqrt{\det (g - \cF)}
      J^{\mu\nu\gamma}\partial_{\gamma}\Theta^{\rho\lambda}
      ([f_{\rho}*g_{\mu\lambda}*h_{\nu}]
      +[f_{\rho}*g_{\mu}*h_{\nu\lambda}] \nonumber
\\ & & ~~~~~~~~~~~~~~~~~~~~~~~~~~~~~~~~~~~~~~~~~
      -[f_{\mu\lambda}*g_{\nu}*h_{\rho}]
      -[f_{\mu}*g_{\nu\lambda}*h_{\rho}]),
\end{eqnarray}
which we have brought into standard index ordering.

Now we are ready to take 
all remaining terms of (\ref{eq:nonassoc}) into account, i.e.
expressions (\ref{eq:NA1O}), (\ref{eq:NA2Ord3}) and the seventh line
of (\ref{eq:nonassoc}). If we rewrite (\ref{eq:NA1O}) as
\begin{eqnarray}
  \label{eq:NA1Ord2}
& &    -\frac{\Ii}{12}\int d^Dx \sqrt{\det (g - \cF)}
      J^{\mu\nu\gamma}\partial_{\gamma}\Theta^{\rho\lambda}
      ([f_{\rho}*g_{\mu\lambda}*h_{\nu}]
      +\frac{1}{2}[f_{\rho}*g_{\mu}*h_{\nu\lambda}] \nonumber
\\ & & ~~~~~~~~~~~~~~~~~~~~~~~~~~~~~~~~~~~~~~~~~
      -\frac{1}{2}[f_{\mu\lambda}*g_{\nu}*h_{\rho}]
      -[f_{\mu}*g_{\nu\lambda}*h_{\rho}]),
\end{eqnarray}
and add (\ref{eq:NA2Ord3}) we obtain 
\begin{eqnarray}
  \label{eq:NA2Ord4}
& &   -\frac{\Ii}{24}\int d^Dx \sqrt{\det (g - \cF)}
      J^{\mu\nu\gamma}\partial_{\gamma}\Theta^{\rho\lambda}
      ([f_{\lambda}*g_{\nu\rho}*h_{\mu}]
      +[f_{\mu}*g_{\nu\rho}*h_{\lambda}]).
\end{eqnarray}
But this expression cancels exactly the next to last line in
(\ref{eq:nonassoc}). 
So we have finally shown that eq. (\ref{eq:assocprod}) is
fulfilled in second derivative order, i.e.
\begin{eqnarray}
 \label{eq:assoczero}
  \int_x (f \circ g)\circ h - f \circ (g \circ h) ~\approx~
  \cO(\partial^3) \;.
\end{eqnarray} 
In particular, we observe that the constant $C_2$ remains undetermined.

\noindent\emph{Ordinary product up to surface terms}

We proceed in checking whether the product (\ref{eq:product}) reduces
to the ordinary product under the integral. This task is greatly simplified by 
observing that all terms with third or higher powers in $\Theta$ can be 
pushed to third derivative order. The linear $\Theta$ term was already shown 
to vanish by the background equation (\ref{eq:eom}) in 
\cite{Herbst:2001ai}, so that it remains to consider
\begin{equation}
 \label{eq:ord2nd} 
 \int f\circ g \approx
 \int \Bigl(f \cdot g - 
            \frac{1}{8} \Th^{\mu\rho} \Th^{\nu\sigma}
            f_{\mu\nu} \cdot g_{\rho\sigma}
   -\frac{1}{12}\Th^{\mu\gamma}\partial_{\gamma}\Th^{\nu\rho}
   (f_{\mu\nu} \cdot g_{\rho}+f_{\rho} \cdot g_{\mu\nu}) 
   +\frac{A}{4}\;\6_\d\Th^{\m\g}\6_\g\Th^{\n\d}\;f_\m*g_\n \Bigr) \;.
\end{equation}
By the usual arguments expression (\ref{eq:ord2nd}) can be rewritten as
\begin{equation}
  \label{eq:ord2nd1}
 \int \Bigl(f \cdot g
          + \frac{(6A-1)}{24} \partial_\sigma \Th^{\mu\rho}
                         \partial_\rho   \Th^{\nu\sigma}
                         f_\mu \cdot g_\nu
          + \frac{1}{24} \Th^{\mu\rho}\Th^{\nu\sigma}
                         \partial_\rho \partial_\sigma 
                         \bigl( \ln \sqrt{\det(g-\cF)} \bigr)~
                         f_\mu \cdot g_\nu
          \Bigr) \;.
\end{equation}
Demanding that expression (\ref{eq:ord2nd1}) becomes the ordinary
product of functions requires $A=\frac{1}{6}$; moreover, it
shows that we have forgotten a contribution to the product, which
is capable of compensating the last term in (\ref{eq:ord2nd1}). In fact, we
involved only tree level and loop diagrams in the product
(\ref{eq:product}), which can be constructed with $\Th$. In particular,
the second term in (\ref{eq:ord2nd1}) comes from a loop diagram in
Kontsevich's expansion.
However, the last term is not of this type and arises much in the same
manner as the integration measure (cf. \cite{Herbst:2001ai}).
Requiring relation (\ref{eq:ordprod}) therefore determines the
explicit dependence of the product on loop contributions, i.e. it
fixes the constant $A$ and the factor in front of the logarithmic
term.

The product (\ref{eq:product}) therefore becomes
\begin{eqnarray}
 \label{eq:finproduct} 
 f\circ g &\Back =\Back & f * g 
    -\frac{1}{12}\Th^{\mu\gamma}\partial_{\gamma}\Th^{\nu\rho}
       \bigl(f_{\mu\nu}*g_{\rho}+f_{\rho}*g_{\mu\nu}\bigr) \nonumber 
\\
& &       + \frac{1}{24} \partial_\sigma \Th^{\mu\rho}
                         \partial_\rho   \Th^{\nu\sigma}
                         f_\mu * g_\nu
          - \frac{1}{24} \Th^{\mu\rho}\Th^{\nu\sigma}
                         \partial_\rho \partial_\sigma 
                         \bigl( \ln \sqrt{\det(g-\cF)} \bigr)~
                         f_\mu * g_\nu \nonumber
\\
& & -\frac{\Ii}{8} \Th^{\mu\gamma}\partial_{\gamma}\Th^{\nu\delta}
  \partial_{\delta}\Th^{\rho\lambda}~
  \Bigl(\Bigl(C_2-\frac{1}{3}\Bigr)\ f_{\mu\rho}*g_{\nu\lambda}
   + C_2\ f_{\nu\lambda}*g_{\mu\rho}\Bigr) \nonumber
\\ 
& & -\frac{\Ii}{48}\Th^{\mu\gamma}\Th^{\nu\delta}
     \partial_{\gamma}\partial_{\delta}\Th^{\rho\lambda}~
     \bigl(f_{\mu\nu\rho}*g_{\lambda} - f_{\lambda}*g_{\mu\nu\rho}\bigr) 
     \nonumber 
\\
& & \frac{1}{2}\frac{1}{12^2}
      (\Th^{\mu\gamma}\partial_{\gamma}\Th^{\rho\lambda})
      (\Th^{\nu\delta}\partial_{\delta}\Th^{\sigma\tau})
      \bigl(f_{\mu\rho\nu\sigma}*g_{\lambda\tau} 
       +2 f_{\mu\rho\tau}*g_{\lambda\nu\sigma}
       + f_{\lambda\tau}*g_{\mu\rho\nu\sigma}\bigr)
\;.
\end{eqnarray}
The coefficient for the loop diagram, $\partial\Theta\partial\Theta$,
coincides with the result of ref. \cite{Dito:2002dr}, whereas the
term $\ln{\sqrt{\det(g-\cF)}}$ represents a new contribution to the product.
One may wonder whether this term spoils relation (\ref{eq:assoczero}),
but it can be eliminated by a gauge transformation (\ref{GaugeTrafo})
and thus has no effect on the associativity.

\section{Conclusion}

We have constructed a non-associative product that is cyclic with
respect to the Born--Infeld measure through second order in the derivative
expansion. To this end we have
evaluated the associator for the product of three 
functions on the world-volume of a curved D-brane, whose consistency with
the topological limit yields the weights for an infinite number of 
Kontsevich graphs as a by-product (cf. fig. 1).
Our product reproduces the Kontsevich formula, including the gauge term, 
but has an additional contribution with a logarithmic derivative of the 
measure that may diverge in the topological limit (note that a vanishing
divergence of the Poisson structure for some measure is required by cyclicity
already in the associative case \cite{Felder:2000nc}). In the context of 
effective low energy actions for open strings in background fields cyclicity, 
rather than associativity, therefore seems to be the crucial property.

We conjecture that our results can be extended to arbitrary orders in the 
derivative expansion, provided that one takes into 
account corrections from vacuum loops to the Born--Infeld measure,
see for instance~\cite{Okawa:1999cm,Andreev:2001xx,Wyllard:2000qe,Fotopoulos:2001pt,Das:2001xy,Wyllard:2001ye,Pal:2001xp}. 
It is well 
known that certain ambiguities exist in the computation of the renormalized 
partition function~\cite{Andreev:2001xx}, which are related to the scheme 
dependence of the renormalization procedure. Some of these ambiguities may be 
fixed by imposing the cyclicity condition using the open string 
partition function as measure. 

Since the non-associativity in the non-topological 
situation comes from the singularities of the boundary OPEs, which we 
removed in~\cite{Herbst:2001ai} by subtraction, a proof of our
conjecture may require an analysis of the Ward  
identities and of the $A_\infty$ structure of open string field
theory~\cite{Gaberdiel:1997ia,Kajiura:2003ax}.

Recently, string-inspired superspace deformations have attracted a lot of
interest~\cite{deBoer:2003dn,Ooguri:2003qp,Ooguri:2003tt,Seiberg:2003yz,Berkovits:2003kj}. Such a deformation arises from considering open superstrings in a
graviphoton background and can be directly calculated using a covariant 
quantum description of superstrings with space-time 
supercoordinates~\cite{Berkovits:2000fe,Berkovits:2002zk,Grassi:2001ug}. 
Clearly, the starting point for these investigations is constant background 
fields. A corresponding investigation of non-constant backgrounds is lacking 
at present. It would be interesting to see how Kontsevich's formula 
generalizes to a non-commutative product on superspace and whether 
non-associativity is constrained in these cases by supersymmetry. Furthermore, 
it would be rewarding to explore the physical aspects of curved brane
geometries, such as brane stabilization due to non-trivial background 
fluxes~\cite{Bachas:2000ik,Alekseev:2000fd,Kling:2000dy,Takayanagi:2001gu} in 
a supersymmetric setting. 

{\it Acknowlegements.} We would like to thank A. Alekseev, G. Felder, 
H. Grosse and K.-G. Schlesinger for helpful discussions.
This work was supported in part by the city of Vienna under grant number 
H-85/2001 and by the Austrian Research Fund FWF under grant number P15553.
The work of A.K. was supported by the DFG 
priority program `String Theory' (SPP 1096).

\appendix

\section{Evaluation of the associator}
\label{sec:2ndordcalc}

In order to compute the associator for a product that is compatible
with the topological limit, we start with the ansatz 
\begin{eqnarray}
  \label{eq:ansatz}
  f\circ g &\Back =\Back & f * g 
    -\frac{1}{12}\Th^{\mu\gamma}\partial_{\gamma}\Th^{\nu\rho}
       (f_{\mu\nu}*g_{\rho}+f_{\rho}*g_{\mu\nu}) \nonumber 
\\
& & -\frac{\Ii}{8}
  \Th^{\mu\gamma}\Th^{\nu\delta}
     \partial_{\gamma}\partial_{\delta}\Th^{\rho\lambda}
        (B_1 f_{\mu\nu\rho}*g_{\lambda} + B_2 f_{\lambda}*g_{\mu\nu\rho} 
         + B_3 f_{\mu\rho}*g_{\nu\lambda}) \\
       & & -\frac{\Ii}{8}
  \Th^{\mu\gamma}\partial_{\gamma}\Th^{\nu\delta}
     \partial_{\delta}\Th^{\rho\lambda}
        (C_1 f_{\mu\rho}*g_{\nu\lambda}+ C_2 f_{\nu\lambda}*g_{\mu\rho}
         + C_3 f_{\mu\nu\rho}*g_{\lambda}+ C_4 f_{\lambda}*g_{\mu\nu\rho}) 
          \nonumber \\
       & & +\frac{1}{16}(\Th^{\mu\gamma}\partial_{\gamma}\Th^{\rho\lambda})
                        (\Th^{\nu\delta}\partial_{\delta}\Th^{\sigma\tau})
         (D_1 f_{\mu\rho\nu\sigma}*g_{\lambda\tau} 
          + D_2 f_{\mu\rho\tau}*g_{\lambda\nu\sigma}
          + D_3 f_{\lambda\tau}*g_{\mu\rho\nu\sigma}), \nonumber
\end{eqnarray}
where we used the notation $\partial_{\mu} f=f_{\mu}$ for the 
derivatives acting on the inserted functions. The coefficients $B_i$, $C_i$ 
and $D_i$ are arbitrary constants and we dropped the gauge term with 
coefficient $A$ of~(\ref{eq:Kgraphs}).
Compatibility with the case of a Poisson manifold 
implies that the associator of three functions 
\begin{equation}
  \label{eq:associator}
  (f \circ g)\circ h - f \circ (g \circ h) = 0 + \cO(J,\partial J)
\end{equation}
only contains terms that are proportional to the Jacobiator 
$J^{\mu\rho\delta}$ (\ref{eq:jacobiator}) or derivatives thereof.

Obviously the terms involving four $\Theta$'s do not mix with 
the other terms in the second derivative order. Inserting the different 
contibutions to the generalized star product into the associator 
(\ref{eq:associator}), we obtain the following $(\Theta\partial\Theta)^2$ 
terms from expanding the lowest order part:
\begin{equation}
  \label{eq:FourThetas1}
  (f * g) * h - f *(g * h) = \frac{1}{32}
     (\Theta^{\mu\gamma}\partial_{\gamma}\Theta^{\rho\lambda})
     (\Theta^{\nu\delta}\partial_{\delta}\Theta^{\sigma\tau})
     \left( [f_{\rho\sigma} * g_{\lambda\tau} * h_{\mu\nu}]
      -[f_{\mu\nu} * g_{\rho\sigma} * h_{\lambda\tau}]\right),
\end{equation}
where square brackets around the product of three or more functions indicate 
that there are no derivatives on $\Theta$'s contained in theses expressions. 
Introducing the notation
\[
f\circ_1 g = -\frac{1}{12}\Th^{\mu\gamma}\partial_{\gamma}\Th^{\nu\rho}
       (f_{\mu\nu}*g_{\rho}+f_{\rho}*g_{\mu\nu})
\]
for the first derivative order we obtain the following terms
\begin{eqnarray}
  \label{eq:FourThetas2}
 (f\circ_1 g) * h - f *(g\circ_1  h) &\Back = \Back & -\frac{1}{48}
     (\Theta^{\mu\gamma}\partial_{\gamma}\Theta^{\rho\lambda})
     (\Theta^{\nu\delta}\partial_{\delta}\Theta^{\sigma\tau})
     ( [f_{\mu\rho\sigma} * g_{\lambda\tau} * h_{\nu}]
      +[f_{\lambda\sigma} * g_{\mu\rho\tau} * h_{\nu}] \nonumber
\\  &\back  \back & 
      +[f_{\nu} * g_{\mu\rho\sigma} * h_{\lambda\tau}]
      +[f_{\nu} * g_{\lambda\sigma} * h_{\mu\rho\tau}]) \nonumber
\\
 (f * g) \circ_1 h - f \circ_1 (g * h) &\Back = \Back & -\frac{1}{48}
     (\Theta^{\mu\gamma}\partial_{\gamma}\Theta^{\rho\lambda})
     (\Theta^{\nu\delta}\partial_{\delta}\Theta^{\sigma\tau})
     ( [f_{\mu\rho\sigma} * g_{\tau} * h_{\nu\lambda}]
      +[f_{\mu\sigma} * g_{\rho\tau} * h_{\nu\lambda}] \nonumber
\\  &\back  \back & 
      +[f_{\rho\sigma} * g_{\mu\tau} * h_{\nu\lambda}]
      +[f_{\sigma} * g_{\mu\rho\tau} * h_{\nu\lambda}]
      +[f_{\lambda\sigma} * g_{\tau} * h_{\mu\nu\rho}] \nonumber
\\  &\back  \back & 
      +[f_{\sigma} * g_{\lambda\tau} * h_{\mu\nu\rho}]
      +[f_{\mu\nu\rho} * g_{\lambda\sigma} * h_{\tau}]
      +[f_{\mu\nu\rho} * g_{\sigma} * h_{\lambda\tau}] \nonumber
\\  &\back  \back & 
      +[f_{\nu\lambda} * g_{\mu\rho\sigma} * h_{\tau}]
      +[f_{\nu\lambda} * g_{\mu\sigma} * h_{\rho\tau}]
      +[f_{\nu\lambda} * g_{\rho\sigma} * h_{\mu\tau}] \nonumber
\\  &\back  \back & 
      +[f_{\nu\lambda} * g_{\sigma} * h_{\mu\rho\tau}]) \nonumber
\\
  (f \circ_1 g) \circ_1 h - f \circ_1 (g \circ_1 h) &\Back = \Back &
 (\frac{1}{12})^2 (\Theta^{\mu\gamma}\partial_{\gamma}\Theta^{\rho\lambda})
     (\Theta^{\nu\delta}\partial_{\delta}\Theta^{\sigma\tau})
     ([f_{\mu\nu\rho\sigma} * g_{\tau} * h_{\lambda}] 
\\  &\back  \back & 
      +[f_{\mu\nu\sigma} * g_{\rho\tau} * h_{\lambda}]
      +[f_{\rho\nu\sigma} * g_{\mu\tau} * h_{\lambda}]
      +[f_{\mu\rho\tau} * g_{\nu\sigma} * h_{\lambda}] \nonumber
\\  &\back  \back &      
      +[f_{\mu\tau} * g_{\rho\nu\sigma} * h_{\lambda}]
      +[f_{\rho\tau} * g_{\mu\nu\sigma} * h_{\lambda}] 
      +[f_{\nu\lambda\sigma} * g_{\tau} * h_{\mu\rho}] \nonumber
\\  &\back  \back & 
      +[f_{\lambda\tau} * g_{\nu\sigma} * h_{\mu\rho}]
      -[f_{\mu\rho} * g_{\nu\sigma} * h_{\lambda\tau}]
      -[f_{\mu\rho} * g_{\tau} * h_{\lambda\nu\sigma}] \nonumber
\\  &\back  \back & 
      -[f_{\lambda} * g_{\mu\nu\sigma} * h_{\rho\tau}]
      -[f_{\lambda} * g_{\rho\nu\sigma} * h_{\mu\tau}] 
      -[f_{\lambda} * g_{\nu\sigma} * h_{\mu\rho\tau}] \nonumber
\\  &\back  \back & 
      -[f_{\lambda} * g_{\mu\tau} * h_{\rho\nu\sigma}] 
      -[f_{\lambda} * g_{\rho\tau} * h_{\mu\nu\sigma}]
      -[f_{\lambda} * g_{\tau} * h_{\mu\nu\rho\sigma}]), \nonumber
\end{eqnarray}
where we have used symmetry properties to cancel some contributions. 
Next, we have to consider the contributions of the second derivative order in
(\ref{eq:ansatz}). We use the following notation:
\[
f \circ_{(\Theta\partial\Theta)^2} g = 
  -\frac{1}{16}(\Th^{\mu\gamma}\partial_{\gamma}\Th^{\rho\lambda})
               (\Th^{\nu\delta}\partial_{\delta}\Th^{\sigma\tau})
               (D_1 f_{\mu\rho\nu\sigma}*g_{\lambda\tau} 
                + D_2 f_{\mu\rho\tau}*g_{\lambda\nu\sigma}
                + D_3 f_{\lambda\tau}*g_{\mu\rho\nu\sigma}).
\]
The terms arising from this contribution are the only ones where the 
arbitrary constants $D_i$ enter the calculations. We obtain
\begin{eqnarray}
  \label{eq:4Theta2}
  (f\circ_{(\Theta\partial\Theta)^2}g)*h &\Back - \Back &
  f*(g\circ_{(\Theta\partial\Theta)^2}h) =  
     \frac{1}{16}(\Th^{\mu\gamma}\partial_{\gamma}\Th^{\rho\lambda})
               (\Th^{\nu\delta}\partial_{\delta}\Th^{\sigma\tau})\times 
\\  &\back  \back & 
  (D_1 [f_{\mu\rho\nu\sigma}*g_{\lambda\tau}*h] 
   + D_2 [f_{\mu\rho\tau}*g_{\lambda\nu\sigma}*h]
   + D_3 [f_{\lambda\tau}*g_{\mu\rho\nu\sigma}*h] \nonumber
\\  &\back  \back & 
   - D_1 [f*g_{\mu\rho\nu\sigma}*h_{\lambda\tau}] 
   - D_2 [f*g_{\mu\rho\tau}*h_{\lambda\nu\sigma}]
   - D_3 [f*g_{\lambda\tau}*h_{\mu\rho\nu\sigma}])\nonumber 
\end{eqnarray}
and
\begin{eqnarray}
  \label{eq:4Theta3}
  (f*g)\circ_{(\Theta\partial\Theta)^2}h &\Back - \Back &
  f\circ_{(\Theta\partial\Theta)^2}(g*h) =  
     \frac{1}{16}(\Th^{\mu\gamma}\partial_{\gamma}\Th^{\rho\lambda})\times
\\  &\back  \back & 
   (D_1 [(f*g)_{[\mu\rho\nu\sigma]}*h_{\lambda\tau}] 
   + D_2 [(f*g)_{[\mu\rho\tau]}*h_{\lambda\nu\sigma}]
   + D_3 [(f*g)_{\lambda\tau}*h_{\mu\rho\nu\sigma}] \nonumber
\\  &\back  \back & 
   - D_1 [f_{\mu\rho\nu\sigma}*(g*h)_{[\lambda\tau]}] 
   - D_2 [f_{\mu\rho\tau}*(g*h)_{[\lambda\nu\sigma]}]
   - D_3 [f_{\lambda\tau}*(g*h)_{[\mu\rho\nu\sigma]}]),\nonumber 
\end{eqnarray}
where the indices in square brackets remind us that the derivatives act only 
on the inserted functions but not on the `star', since these terms are 
already of second derivative order. In expanding these expressions we have 
to be careful, because of the symmetries mentioned above. Putting 
(\ref{eq:4Theta2}) and (\ref{eq:4Theta3}) together we find that all terms 
containing undifferentiated functions cancel. 
Comparing the result with (\ref{eq:FourThetas1}) and (\ref{eq:FourThetas2}), 
we observe that there are only two contributions with four derivatives 
acting on $f$ and two contributions with four derivatives acting on $g$. From 
the index structure, antisymmetrization makes it clear that terms 
containing four derivatives acting on the same inserted function can never be 
absorbed into a term proportional to a Jacobiator. Thus these terms have to 
cancel
\[
(\Th^{\mu\gamma}\partial_{\gamma}\Th^{\rho\lambda})
(\Th^{\nu\delta}\partial_{\delta}\Th^{\sigma\tau})~
[f_{\mu\rho\nu\sigma}*g_{\lambda}*h_{\tau}]~
\biggl(\Bigl(\frac{1}{12}\Bigr)^2-\frac{D_1}{8}\biggr)=0,
\]
which fixes $D_1$ to be $D_1=1/18$. In the same way we obtain
\[
D_3=D_1=\frac{1}{18}.
\]
Now the remaining constant $D_2$ has to be chosen in such a way that all 
terms combine to expressions proportional to a Jacobiator. Let us collect all 
terms of the form $[\partial^3 f*\partial g * \partial^2 h]$. After 
rearranging the indices, and using the symmetries, we find
\[
\Bigl(\frac{1}{9}-D_2\Bigr)~[f_{\mu\rho\tau}*g_{\lambda}*h_{\nu\sigma}] + 
\Bigl(\frac{1}{9}+D_2\Bigr)~[f_{\mu\rho\sigma}*g_{\nu}*h_{\lambda\tau}] +
\frac{2}{9}~[f_{\mu\rho\nu}*g_{\tau}*h_{\lambda\sigma}] +
\frac{2}{9}~[f_{\mu\rho\tau}*g_{\sigma}*h_{\lambda\nu}].
\]
The first term has to vanish, since it cannot, because of the index
structure, be expressed as part of a Jacobiator. This fixes the
remaining constant to be 
\[
D_2 = \frac{1}{9}.
\]
With the same value the remaining three terms are cyclic in $\nu\sigma\tau$ 
and thus turn the prefactor 
$(\Th^{\nu\delta}\partial_{\delta}\Th^{\sigma\tau})$ into a full Jacobiator
\begin{equation}
  \label{eq:J1}
\Bigl(\frac{1}{12}\Bigr)^2 (\Th^{\mu\gamma}\partial_{\gamma}\Th^{\rho\lambda})
        J^{\nu\sigma\tau}\, 2\, [f_{\mu\rho\sigma}*g_{\nu}*h_{\lambda\tau}],  
\end{equation}
where we have written the full expression with the correct numerical 
prefactor. Repeating this procedure for the other terms gives the 
following contribution to the associator
\begin{eqnarray}
  \label{eq:assoz1}
  (f \circ g)\circ h & \Back - \Back & f \circ (g \circ h) = 
      2\, \Bigl(\frac{1}{12}\Bigr)^2 (\Th^{\mu\gamma}\partial_{\gamma}\Th^{\rho\lambda})
        J^{\nu\sigma\tau}\, \times
\\    & \Back  \Back &  
    \Bigl( [f_{\mu\rho\nu}*g_{\tau}*h_{\lambda\sigma}]  
    + [f_{\mu\rho\nu}*g_{\lambda\tau}*h_{\sigma}]
    + [f_{\nu\lambda}*g_{\mu\rho\tau}*h_{\sigma}] \nonumber
\\    & \back  \back &
    + [f_{\nu}*g_{\mu\rho\tau}*h_{\lambda\sigma}] 
    +2\, [f_{\mu\nu}*g_{\rho\tau}*h_{\lambda\sigma}]
    +2\, [f_{\lambda\nu}*g_{\rho\tau}*h_{\mu\sigma}] \nonumber
\\    & \back  \back &
    + [f_{\nu}*g_{\lambda\tau}*h_{\mu\rho\sigma}] 
    + [f_{\nu\lambda}*g_{\tau}*h_{\mu\rho\sigma}]\Bigr).  \nonumber
\end{eqnarray}
Now we turn to the next contributions arising from our ansatz 
(\ref{eq:ansatz}). Let us consider the part proportional to 
$\Theta\Theta\partial\partial\Theta$ and collect all terms in the 
associator that arise from expanding lower order parts of the generalized 
star product 
\begin{eqnarray}
  \label{eq:66Theta1}
  (f * g) * h - f *(g * h) & \Back = \Back & \frac{\Ii}{16}
  \Theta^{\mu\gamma}\Theta^{\nu\delta}
  \partial_{\gamma}\partial_{\delta}\Theta^{\rho\lambda}
  \Bigl( [f_{\mu\nu}*g_{\rho}*h_{\lambda}] - 
  [f_{\rho}*g_{\lambda}*h_{\mu\nu}]\Bigr)
\\
  (f \circ_1 g) * h - f *(g \circ_1 h) & \Back = \Back & \frac{\Ii}{24} 
  \Theta^{\mu\gamma}\Theta^{\nu\delta}
  \partial_{\gamma}\partial_{\delta}\Theta^{\rho\lambda}
  \Bigl( [f_{\nu\rho}*g_{\lambda}*h_{\mu}] \nonumber
\\
 & \Back  \Back &
   + [f_{\lambda}*g_{\nu\rho}*h_{\mu}]
   + [f_{\mu}*g_{\nu\rho}*h_{\lambda}]
   + [f_{\mu}*g_{\lambda}*h_{\nu\rho}]\Bigr). \nonumber
\end{eqnarray}
From the contributions at second derivative order, terms with 
undifferentiated functions again do not survive, and we obtain
\begin{eqnarray}
  \label{eq:66Theta2}
  (f \circ_{\partial\partial\Theta} g) * h
  &\Back  \Back & + (f * g) \circ_{\partial\partial\Theta} h 
  - f *(g \circ_{\partial\partial\Theta} h)
  - f \circ_{\partial\partial\Theta} (g * h) = \nonumber
\\ 
  - \frac{\Ii}{8} \Theta^{\mu\gamma}\Theta^{\nu\delta}
  \partial_{\gamma}\partial_{\delta}\Theta^{\rho\lambda} &\Back  \Back &
  \Bigl( C_7~ ([f_{\mu\nu}*g_{\rho}*h_{\lambda}]
          +2~ [f_{\mu\rho}*g_{\nu}*h_{\lambda}]
          + [f_{\rho}*g_{\mu\nu}*h_{\lambda}]
          +2~ [f_{\nu}*g_{\mu\rho}*h_{\lambda}]) \nonumber
\\
   &\Back  \Back & - C_8~ ([f_{\lambda}*g_{\mu\nu}*h_{\rho}]
          +2~ [f_{\lambda}*g_{\mu\rho}*h_{\nu}]
          + [f_{\lambda}*g_{\rho}*h_{\mu\nu}]
          +2~ [f_{\lambda}*g_{\nu}*h_{\mu\rho}]) \nonumber
\\
   &\Back  \Back & + C_9~ ([f_{\mu}*g_{\rho}*h_{\nu\lambda}]
          + [f_{\rho}*g_{\mu}*h_{\nu\lambda}]
          - [f_{\mu\rho}*g_{\nu}*h_{\lambda}]
          - [f_{\mu\rho}*g_{\lambda}*h_{\nu}])\Bigr). \nonumber
\end{eqnarray}
Collecting the terms with two derivatives acting on the insertion $f$, we 
find
\begin{eqnarray}
  \label{eq:C7}
  \frac{\Ii}{16} \Theta^{\mu\gamma}\Theta^{\nu\delta}
  \partial_{\gamma}\partial_{\delta}\Theta^{\rho\lambda} &\Back  \Back &
  \Bigl((1-2B_1)~[f_{\mu\nu}*g_{\rho}*h_{\lambda}]
  +\Bigl(\frac{2}{3}+2 B_3\Bigr)~[f_{\mu\rho}*g_{\lambda}*h_{\nu}] \nonumber
\\ &\Back  \Back &
  +(4 B_1-2 B_3)~ [f_{\mu\lambda}*g_{\nu}*h_{\rho}]\Bigr).
\end{eqnarray}
We observe that the three terms are cyclic in $\nu\rho\lambda$,
provided the coefficients are equal; this fixes the constants to be
\[
B_1=\frac{1}{6}, \quad  B_3=0.
\]
Going through the same procedure for the terms with two derivatives 
acting on $g$, we find 
\begin{eqnarray}
  \label{eq:C8}
  \frac{\Ii}{16} \Theta^{\mu\gamma}\Theta^{\nu\delta}
  \partial_{\gamma}\partial_{\delta}\Theta^{\rho\lambda} &\Back  \Back &
  \Bigl((-1-2B_2)~[f_{\rho}*g_{\lambda}*h_{\mu\nu}]
  -\Bigl(\frac{2}{3}+2 B_3\Bigr)~[f_{\nu}*g_{\rho}*h_{\mu\lambda}] \nonumber
\\ &\Back  \Back &
  +(4 B_2-2 B_3) [f_{\lambda}*g_{\nu}*h_{\mu\rho}]\Bigr),
\end{eqnarray}
which fixes the constants to
\[
B_2=-\frac{1}{6}, \quad B_3=0,
\]
and is thus compatible with the above values. With these values for $B_1$ 
and $B_2$ the remaining terms cancel and we are left with the following 
result
\begin{eqnarray}
  \label{eq:66Theta3}
  \frac{\Ii}{24}\Theta^{\mu\gamma}\Theta^{\nu\delta}
  \partial_{\gamma}\partial_{\delta}\Theta^{\rho\lambda} &\Back  \Back &
  \Bigl([f_{\mu\nu}*g_{\rho}*h_{\lambda}]
  +[f_{\mu\rho}*g_{\lambda}*h_{\nu}]
  +[f_{\mu\lambda}*g_{\nu}*h_{\rho}] \nonumber 
\\ &\Back  \Back &
  -[f_{\rho}*g_{\lambda}*h_{\mu\nu}]
  -[f_{\nu}*g_{\rho}*h_{\mu\lambda}]
  -[f_{\lambda}*g_{\nu}*h_{\mu\rho}]\Bigr).
\end{eqnarray}
To turn these expressions into terms proportional to a Jacobiator, we 
rewrite
\[
\Theta^{\mu\gamma}\Theta^{\nu\delta}
  \partial_{\gamma}\partial_{\delta}\Theta^{\rho\lambda} = 
\Theta^{\mu\gamma}\partial_{\gamma}(\Theta^{\nu\delta}
  \partial_{\delta}\Theta^{\rho\lambda}) - 
\Theta^{\mu\gamma}\partial_{\gamma}\Theta^{\nu\delta}
  \partial_{\delta}\Theta^{\rho\lambda}.
\]
Then the first term on the right-hand side gives rise to
\begin{equation}
  \label{eq:6J}
  \frac{\Ii}{24}\Theta^{\mu\gamma}\partial_{\gamma}J^{\nu\rho\lambda} 
  \bigl([f_{\mu\nu}*g_{\rho}*h_{\lambda}]-
  [f_{\rho}*g_{\lambda}*h_{\mu\nu}] \bigr).
\end{equation}
The remaining terms proportional to $\Theta\partial\Theta\partial\Theta$,
\begin{eqnarray}
  \label{eq:rest}
 - \frac{\Ii}{24} \Theta^{\mu\gamma}\partial_{\gamma}\Theta^{\nu\delta}
  \partial_{\delta}\Theta^{\rho\lambda} &\Back  \Back &
  \Bigl([f_{\mu\nu}*g_{\rho}*h_{\lambda}]
  +[f_{\mu\rho}*g_{\lambda}*h_{\nu}]
  +[f_{\mu\lambda}*g_{\nu}*h_{\rho}] \nonumber 
\\ &\Back  \Back &
  -[f_{\rho}*g_{\lambda}*h_{\mu\nu}]
  -[f_{\nu}*g_{\rho}*h_{\mu\lambda}]
  -[f_{\lambda}*g_{\nu}*h_{\mu\rho}]\Bigr),
\end{eqnarray}
still have to be considered. To this end we follow the above procedure and 
collect the terms proportional to $\Theta\partial\Theta\partial\Theta$ arising 
from lower derivative orders:
\begin{eqnarray}
  \label{eq:Th6Th6Th}
  (f\circ_1 g) * h - f *(g\circ_1  h) &\Back = \Back &
  \frac{\Ii}{24} \Theta^{\mu\gamma}\partial_{\gamma}\Theta^{\nu\delta}
  \partial_{\delta}\Theta^{\rho\lambda} \Bigl([f_{\nu\rho}*g_{\lambda}*h_{\mu}] 
  \nonumber
\\ &\Back  \Back &
  +[f_{\lambda}*g_{\nu\rho}*h_{\mu}]
  +[f_{\mu}*g_{\nu\rho}*h_{\lambda}]
  +[f_{\mu}*g_{\lambda}*h_{\nu\rho}]\Bigr) \nonumber
\\
  (f* g) \circ_1 h - f \circ_1(g* h) &\Back = \Back &
  \frac{\Ii}{24} \Theta^{\mu\gamma}\partial_{\gamma}\Theta^{\nu\delta}
  \partial_{\delta}\Theta^{\rho\lambda}\Bigl([f_{\mu\rho}*g_{\lambda}*h_{\nu}] 
  \nonumber
\\ &\Back  \Back &
   +[f_{\rho}*g_{\mu\lambda}*h_{\nu}]
   -[f_{\rho}*g_{\lambda}*h_{\mu\nu}]
   +[f_{\mu\nu}*g_{\rho}*h_{\lambda}]   \nonumber
\\ &\Back  \Back &
   -[f_{\nu}*g_{\mu\rho}*h_{\lambda}]
   -[f_{\nu}*g_{\rho}*h_{\mu\lambda}]\Bigr) \nonumber
\\ &\Back  \Back &
   -\frac{\Ii}{24} \Theta^{\gamma\delta}\partial_{\gamma}\Theta^{\mu\nu}
  \partial_{\delta}\Theta^{\rho\lambda}\Bigl([f_{\mu\rho}*g_{\nu}*h_{\lambda}] 
  \nonumber
\\ &\Back  \Back &
   +[f_{\mu}*g_{\nu\rho}*h_{\lambda}]
   -[f_{\lambda}*g_{\mu\rho}*h_{\nu}]
   -[f_{\lambda}*g_{\mu}*h_{\nu\rho}]\Bigr) 
\end{eqnarray}
Note the different tensorial structures of these terms. Owing to the
symmetries, 
the first and second terms in the last line of (\ref{eq:Th6Th6Th}) cancel. 
We rearrange the two terms in the second equation of (\ref{eq:Th6Th6Th}) 
proportional to $\Theta^{\gamma\delta}\partial_{\gamma}\Theta^{\mu\nu}
\partial_{\delta}\Theta^{\rho\lambda}$ by 
\begin{eqnarray}
  \label{eq:6Th^2}
   -\frac{\Ii}{24} \Theta^{\gamma\delta}\partial_{\gamma}\Theta^{\mu\nu}
  \partial_{\delta}\Theta^{\rho\lambda}& \Back  \Back &
  \Bigl([f_{\mu\rho}*g_{\nu}*h_{\lambda}]-[f_{\lambda}*g_{\mu}*h_{\nu\rho}]\Bigr)=
\\ & \Back \Back &
  \frac{\Ii}{24} J^{\mu\nu\delta}\partial_{\delta}\Theta^{\rho\lambda}
  \Bigl([f_{\mu\rho}*g_{\nu}*h_{\lambda}] - [f_{\lambda}*g_{\mu}*h_{\nu\rho}]\Bigr)
  \nonumber
\\ & \Back \Back &
  -\frac{\Ii}{24} \Theta^{\mu\gamma}\partial_{\gamma}\Theta^{\nu\delta}
   \partial_{\delta}\Theta^{\rho\lambda}\Bigl([f_{\mu\rho}*g_{\nu}*h_{\lambda}] 
    - [f_{\lambda}*g_{\mu}*h_{\nu\rho}] \nonumber
\\ & \Back \Back & ~~~~~~~~~~~~~~~~~~~~~~~~~
    - [f_{\nu\rho}*g_{\mu}*h_{\lambda}] 
    + [f_{\lambda}*g_{\nu}*h_{\mu\rho}]\Bigr). \nonumber
\end{eqnarray}
Putting the pieces of (\ref{eq:rest}), (\ref{eq:Th6Th6Th}) and 
(\ref{eq:6Th^2}) together, we find:
\begin{eqnarray}
  \label{eq:J6Th}
\frac{\Ii}{24}\Theta^{\mu\gamma}\partial_{\gamma}\Theta^{\nu\delta}
   \partial_{\delta}\Theta^{\rho\lambda}&\Back\Back&
       \Bigl([f_{\nu\rho}*g_{\lambda}*h_{\mu}]
       + [f_{\nu\rho}*g_{\mu}*h_{\lambda}]
       + [f_{\lambda}*g_{\nu\rho}*h_{\mu}]
       + [f_{\mu}*g_{\nu\rho}*h_{\lambda}] 
\\ &\Back\Back& 
       + [f_{\rho}*g_{\mu\lambda}*h_{\nu}]
       + [f_{\nu}*g_{\mu\lambda}*h_{\rho}]
       + [f_{\mu}*g_{\lambda}*h_{\nu\rho}]
       + [f_{\lambda}*g_{\mu}*h_{\nu\rho}]\Bigr). \nonumber
\end{eqnarray}
The terms with two derivatives acting on the insertion $g$ can be recast to 
give Jacobiators; thus the contribution of the lower derivative order 
(\ref{eq:rest}) and (\ref{eq:Th6Th6Th}) to the associator is given by
\begin{eqnarray}
  \label{eq:result1}
&& \frac{\Ii}{24}J^{\mu\nu\delta}\partial_{\delta}\Theta^{\rho\lambda}
     \Bigl([f_{\mu\rho}*g_{\nu}*h_{\lambda}] 
     + [f_{\lambda}*g_{\nu}*h_{\mu\rho}]
     + [f_{\mu}*g_{\nu\rho}*h_{\lambda}]
     + [f_{\lambda}*g_{\nu\rho}*h_{\mu}]\Bigr) 
\\
&& +\frac{\Ii}{24}\Theta^{\mu\gamma}\partial_{\gamma}\Theta^{\nu\delta}
   \partial_{\delta}\Theta^{\rho\lambda}
     \Bigl([f_{\nu\rho}*g_{\lambda}*h_{\mu}]
     + [f_{\nu\rho}*g_{\mu}*h_{\lambda}]
     + [f_{\mu}*g_{\lambda}*h_{\nu\rho}]
     + [f_{\lambda}*g_{\mu}*h_{\nu\rho}]\Bigr). \nonumber
\end{eqnarray}
Now we compute the terms that arise from the second derivative order 
contribution to the $\circ$-product proportional to $C_1$ and $C_2$. Again 
terms involving undifferentiated functions cancel and we obtain 
\begin{eqnarray}
  \label{eq:Th6Th6Th2O}
&\back &(f \circ_{\Theta\partial\Theta\partial\Theta} g)*h
  +(f*g)\circ_{\Theta\partial\Theta\partial\Theta} h
  -f*(g\circ_{\Theta\partial\Theta\partial\Theta} h)
  -f \circ_{\Theta\partial\Theta\partial\Theta}(g*h) =
\\
&\back &-\frac{\Ii}{8}\Theta^{\mu\gamma}\partial_{\gamma}\Theta^{\nu\delta}
   \partial_{\delta}\Theta^{\rho\lambda}
   \Bigl(C_1([f_{\mu}*g_{\rho}*h_{\nu\lambda}]
        +[f_{\rho}*g_{\mu}*h_{\nu\lambda}])
    +C_2([f_{\nu}*g_{\lambda}*h_{\mu\rho}]
        +[f_{\lambda}*g_{\nu}*h_{\mu\rho}]) \nonumber
\\ &\back &~~~~~~~~~~~~~~~~~~~~~  
    -C_1([f_{\mu\rho}*g_{\nu}*h_{\lambda}]
        +[f_{\mu\rho}*g_{\lambda}*h_{\nu}])
    -C_2([f_{\nu\lambda}*g_{\mu}*h_{\rho}]
        +[f_{\nu\lambda}*g_{\rho}*h_{\mu}])\Bigr). \nonumber
\end{eqnarray}
In order to arrange the pieces from (\ref{eq:result1}) and
(\ref{eq:Th6Th6Th2O}) in terms of  Jacobiators we have to impose the
condition 
\[
C_2 - C_1 = \frac{1}{3}
\]
on the constants $C_1$ and $C_2$. Eventually we obtain
\begin{eqnarray}
  \label{eq:result2}
  \frac{\Ii}{24}J^{\mu\nu\delta}\partial_{\delta}\Theta^{\rho\lambda} &\back &
  \Bigl(3C_2[f_{\mu\rho}*g_{\nu}*h_{\lambda}]
  +(3C_2-1)[f_{\mu\rho}*g_{\lambda}*h_{\nu}] \nonumber
\\ &\back & ~~~~~~~~
  +[f_{\mu}*g_{\nu\rho}*h_{\lambda}]
  +[f_{\lambda}*g_{\nu\rho}*h_{\mu}] \nonumber
\\ &\back & 
  -3C_2[f_{\mu}*g_{\rho}*h_{\nu\lambda}]
  -(3C_2-1)[f_{\rho}*g_{\mu}*h_{\nu\lambda}]\Bigr).
\end{eqnarray}
This already completes the calculations of the associator 
(\ref{eq:associator}), since the terms proportional to $C_3$ and $C_4$
cannot be recast into Jacobiators, in view of the symmetrization of
$\mu\nu\rho$. Thus the coefficients $C_3$ and $C_4$ have to be zero.
Hence we obtain the final result
\begin{eqnarray}
  \label{eq:finresult}
  (f \circ g)\circ h &\Back -\Back &  f \circ (g \circ h) = 
  \frac{1}{6} J^{\mu\nu\rho}[f_{\mu}*g_{\nu}*h_{\rho}] \nonumber 
\\  & \Back \back & ~~~
  +2\,\Bigl(\frac{1}{12}\Bigr)^2 
    (\Th^{\mu\gamma}\partial_{\gamma}\Th^{\rho\lambda})
        J^{\nu\sigma\tau} \Bigl( [f_{\mu\rho\nu}*g_{\tau}*h_{\lambda\sigma}]
    + [f_{\mu\rho\nu}*g_{\lambda\tau}*h_{\sigma}] \nonumber
\\    & \back  \back & ~~~~~~~~~~~~~~~~~
    + [f_{\nu\lambda}*g_{\mu\rho\tau}*h_{\sigma}] 
    + [f_{\nu}*g_{\mu\rho\tau}*h_{\lambda\sigma}] \nonumber
    +2\, [f_{\mu\nu}*g_{\rho\tau}*h_{\lambda\sigma}]
\\    & \back  \back & ~~~~~~~~~~~~~~~~~ 
    +2\, [f_{\lambda\nu}*g_{\rho\tau}*h_{\mu\sigma}] 
    + [f_{\nu}*g_{\lambda\tau}*h_{\mu\rho\sigma}] 
    + [f_{\nu\lambda}*g_{\tau}*h_{\mu\rho\sigma}]\Bigr)  \nonumber
\\  & \Back  \Back & ~~~
  +\frac{\Ii}{24}J^{\mu\nu\delta}\partial_{\delta}\Theta^{\rho\lambda} 
  \Bigl(3C_2[f_{\mu\rho}*g_{\nu}*h_{\lambda}]
  +(3C_2-1)[f_{\mu\rho}*g_{\lambda}*h_{\nu}] \nonumber
\\  & \Back  \Back & ~~~~~~~~~~~~~~~~~
  -3C_2[f_{\mu}*g_{\rho}*h_{\nu\lambda}]
  -(3C_2-1)[f_{\rho}*g_{\mu}*h_{\nu\lambda}] \nonumber
\\  & \Back  \Back & ~~~~~~~~~~~~~~~~
  +[f_{\mu}*g_{\nu\rho}*h_{\lambda}]
  +[f_{\lambda}*g_{\nu\rho}*h_{\mu}] \Bigr) \nonumber
\\  & \Back  \Back & ~~~
   +\frac{\Ii}{24}\Theta^{\mu\gamma}\partial_{\gamma}J^{\nu\rho\lambda} 
   \Bigl([f_{\mu\nu}*g_{\rho}*h_{\lambda}]-
   [f_{\rho}*g_{\lambda}*h_{\mu\nu}]\Bigr).
\end{eqnarray}
It contains one free parameter, namely $C_2$. The product 
(\ref{eq:ansatz}) then reads
\begin{eqnarray}
 f\circ g &\Back =\Back & f * g 
    -\frac{1}{12}\Th^{\mu\gamma}\partial_{\gamma}\Th^{\nu\rho}
     ~\Bigl(f_{\mu\nu}*g_{\rho}+f_{\rho}*g_{\mu\nu}\Bigr) \nonumber 
\\
& & -\frac{\Ii}{8} \Th^{\mu\gamma}\partial_{\gamma}\Th^{\nu\delta}
  \partial_{\delta}\Th^{\rho\lambda}~
  \Bigl(\Bigl(C_2-\frac{1}{3}\Bigr)\ f_{\mu\rho}*g_{\nu\lambda}
   + C_2\ f_{\nu\lambda}*g_{\mu\rho}\Bigr) \nonumber
\\ 
& & -\frac{\Ii}{48}\Th^{\mu\gamma}\Th^{\nu\delta}
     \partial_{\gamma}\partial_{\delta}\Th^{\rho\lambda}~
     \Bigl(f_{\mu\nu\rho}*g_{\lambda} - 
     f_{\lambda}*g_{\mu\nu\rho}\Bigr) \nonumber 
\\
& & \frac{1}{2}\frac{1}{12^2}
      (\Th^{\mu\gamma}\partial_{\gamma}\Th^{\rho\lambda})
      (\Th^{\nu\delta}\partial_{\delta}\Th^{\sigma\tau})~
      \Bigl(f_{\mu\rho\nu\sigma}*g_{\lambda\tau} 
       +2 f_{\mu\rho\tau}*g_{\lambda\nu\sigma}
       + f_{\lambda\tau}*g_{\mu\rho\nu\sigma}\Bigr), \nonumber
\end{eqnarray}
and the coefficients coincide with those known from 
Kontsevich's formula~\cite{Kontsevich:1997vb,Penkava:1998xx,Zotov:2001ec}.


\end{document}